\documentclass[aps,prl,superscriptaddress,twocolumn,showpacs]{revtex4}
\usepackage{graphics}
\usepackage{epsfig}
\usepackage{color}
\usepackage{stackrel}
\usepackage{amsfonts}
\usepackage{wrapfig}
\usepackage{pstricks}
\usepackage{pst-node}
\usepackage{bm}
\usepackage{dcolumn}% Align table columns on decimal point
\usepackage{multirow}
\usepackage{epstopdf}
\usepackage{subfigure}
\usepackage{amssymb}
\usepackage{amsmath}
\usepackage{commath}
\usepackage{graphicx,bm}
\usepackage{verbatim}
\usepackage{booktabs}
\usepackage[para]{threeparttable}
\usepackage{etoolbox}
\usepackage{lipsum}

\begin{document}
\title{Virtual trions in the photoluminescence of monolayer transition-metal dichalcogenides}

\author{Dinh~Van~Tuan}
%\altaffiliation{These authors contributed equally to this work.}
\affiliation{Department of Electrical and Computer Engineering, University of Rochester, Rochester, New York 14627, USA}

\author{Aaron~M.~Jones}
\affiliation{Department of Physics, University of Washington, Seattle, Washington 98195, USA}

\author{Min~Yang}
%\altaffiliation{These authors contributed equally to this work.}
\affiliation{Department of Electrical and Computer Engineering, University of Rochester, Rochester, New York 14627, USA}

\author{Xiaodong~Xu}
\affiliation{Department of Physics, University of Washington, Seattle, Washington 98195, USA}
\affiliation{Department of Materials Science and Engineering, University of Washington, Seattle, Washington 98195, USA}.

%Department of Physics and Center of Theoretical and Computational Physics, University of Hong Kong, Hong Kong, China. 
%Materials Science and Technology Division, Oak Ridge National Laboratory, Oak Ridge, Tennessee 37831, USA. 
%Department of Materials Science and Engineering, University of Tennessee, Knoxville, Tennessee 37996, USA. 
%Department of Physics and Astronomy, University of Tennessee, Knoxville, Tennessee 37996, USA. 
% Advanced Materials Laboratory, National Institute for Materials Science, Tsukuba, Ibaraki 305-0044, Japan. 

\author{Hanan~Dery}
\altaffiliation{hanan.dery@rochester.edu}
\affiliation{Department of Electrical and Computer Engineering, University of Rochester, Rochester, New York 14627, USA}
\affiliation{Department of Physics and Astronomy, University of Rochester, Rochester, New York 14627, USA}

\begin{abstract}
Photoluminescence experiments from monolayer transition-metal dichalcogenides often show that the binding energy of trions is conspicuously similar to the energy of optical phonons. This enigmatic coincidence calls into question whether phonons are involved in the radiative recombination process. We address this problem, unraveling an intriguing optical transition mechanism. Its initial state is a localized charge (electron or hole) and delocalized exciton. The final state is the localized charge, phonon and photon. In between, the intermediate state of the system is a virtual trion formed when the localized charge captures the exciton through emission of the phonon. We analyze the difference between radiative recombinations that involve real and virtual trions (i.e., with and without a phonon), providing useful ways to distinguish between the two in experiment.
\end{abstract}
\maketitle
%  The phonon energy is close to but not necessarily the exact one needed for transition into the intermediate virtual state, and this resemblance often leads to widespread confusion where the optical transition is attributed to a real trion instead of a phonon-assisted process. We analyze the differences between the two mechanisms, providing useful ways to identify the peaks in the emission spectrum.

The coupling between phonons and charged particles in monolayer transition-metal dichalcogenides (ML-TMDs) exhibits unique behavior due to their atomically thin nature. In addition to a strong Fr\"{o}hlich interaction \cite{Kaasbjerg_PRB12,Sohier_PRB16,Thilagam_JAP16,Xiao_JPCM17},  thickness fluctuations of the ML due to homopolar optical phonons decrease charge mobility \cite{Fivaz_PR67}, while soft undulations of the ML due to acoustic flexural phonons enhance spin relaxation \cite{Cheiwchanchamnangij_PRB13,Song_PRL13}.  In the context of optical properties, phonons induce decoherence, energy relaxation, and mass change of exciton complexes \cite{Song_PRL13,Qiu_PRL13,Kaasbjerg_PRB14,Salehzadeh_NanoLett14,Carvalho_PRL15,Wang_PRL15,Moody_NatComm15,Dery_PRB15,Dey_PRL16,Danovich_IJSTQE,Chow_2DMA,VanTuan_arXiv18, Shree_arXiv18}.  A noticeable property is that the Raman-active phonon energies  nearly resonate with the energy difference between the exciton and trion spectral lines (binding energy of the trion) \cite{Jadczak_PRB17,Jones_NatPhys16}. This energy proximity has often led to widespread confusion  in the assignment of optical transitions in photoluminescence (PL) experiments.

%While this energy proximity can be used for optical refrigeration applications \cite{},  it often leads to a widespread confusion  in the assignment of optical transitions in photoluminescence (PL) experiments.
%\cite{Splendiani_NanoLett10,Mak_PRL10,Xiao_PRL12,Xu_NatPhys14,VanTuan_PRX17,Wang_RMP18}

\begin{figure*}
\includegraphics[width=18cm]{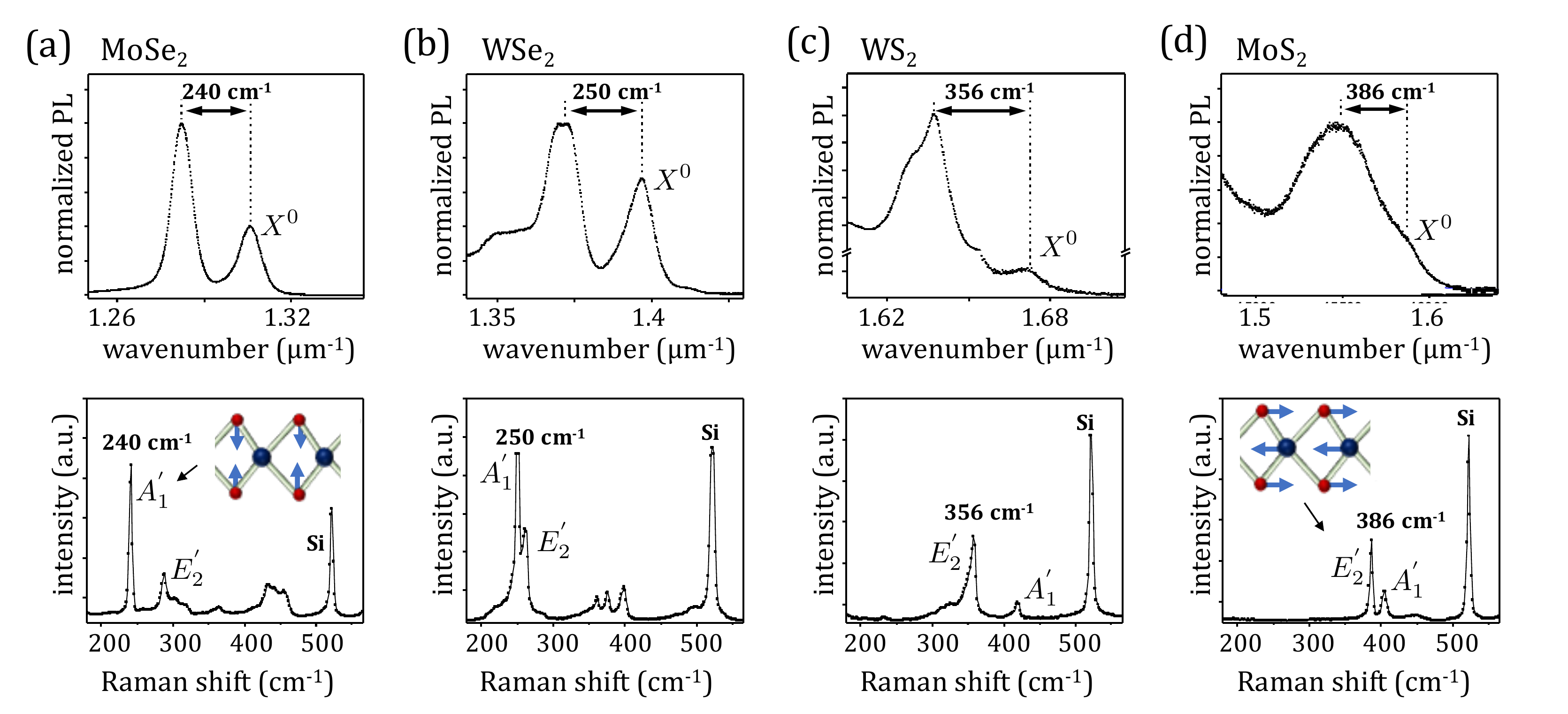}
 \caption{ (a)-(d) Normalized PL (top) and Raman spectra (bottom) of common ML-TMDs supported on Si/SiO$_2$ substrates. The normalized PL includes the neutral exciton peak ($X^0$) and a peak that is commonly associated with the  negatively-charged trion in $n$-type samples. The energy difference between these peaks matches the energy of the dominant Raman active phonon mode in each of the MLs, as is readily seen in the top and bottom panels. The active Raman phonons are the homopolar and longitudinal-optical modes, denoted by $A_{1}^{'}$ and $E_{2}^{'}$, respectively. Their atomic displacements are indicated in the insets of (a) and (d). The Raman spectra also show the active optical phonon mode of the Si substrate around 520~cm$^{-1}$.} \label{fig:exp}
\end{figure*}

We explain the confusion by inspecting the experimental data in Fig.~\ref{fig:exp}. The top panels show the normalized PL from MoSe$_2$, WSe$_2$, WS$_2$, and   MoS$_2$ (left to right), where all the MLs are unintentionally $n$-type and all are supported on Si/SiO$_2$ substrates. For details on the experiment, see Ref.~\cite{Jones_NatNano13}. In each case, the emission spectra have two peaks where the higher energy one is attributed to neutral excitons ($X^0$).  Typically, it is tempting to attribute the low-energy peak to negative trions because some of these peaks also appear in the absorption spectra when electrons are added to the ML through a gate voltage \cite{Wang_NanoLett17,Courtade_PRB17,VanTuan_PRX17}.  However, when performing Raman spectroscopy on the same samples (lower panels of Fig.~\ref{fig:exp}),  we find that the energy of the dominant Raman-active phonon mode matches the energy difference between $X^0$ and the low-energy peak in the PL. This systematic behavior raises the question: Does the low energy peak in the PL stem from recombination of real trions or is it phonon-assisted recombination of excitons? This important question is addressed in this Letter.

Phonon-assisted optical transitions of neutral excitons, governed by the Fr\"{o}hlich interaction \cite{Frohlich_AP54,Feynman_PR55,Cardona_book,Shields_PRB92}, are weak  in ML-TMDs. It is a result of the charge neutrality and small size of the exciton as well as the similar effective masses of electrons and holes. Combined together,  the interactions of the electron and hole with the phonon-induced macroscopic electric field cancel out \cite{VanTuan_arXiv18}. We offer an alternative phonon-assisted recombination scenario facilitated by virtual trion states. Here, the strong Fr\"{o}hlich coupling between a localized electron (or hole) and the lattice is used to capture a nearby exciton by emission of a phonon. The capture mechanism is considered virtual during radiative recombination if $\Delta E = E_T + E_{\lambda} - E_X - E_{\ell} \neq 0$, where the energies are of the localized trion ($E_T$), exciton ($E_X$), localized charge ($E_{\ell}$), and  phonon ($E_{\lambda}$). Energy conservation dictates that the exciton is converted to a phonon plus photon.  Furthermore, if $\Delta E$ is close to zero, the emission spectrum includes a single dominant phonon-assisted optical transition  instead of a series of phonon-replicas that decay according to the standard Huang-Rhys analysis \cite{Huang_PRS50}.

\begin{figure}[b]
\includegraphics[width=8.5cm]{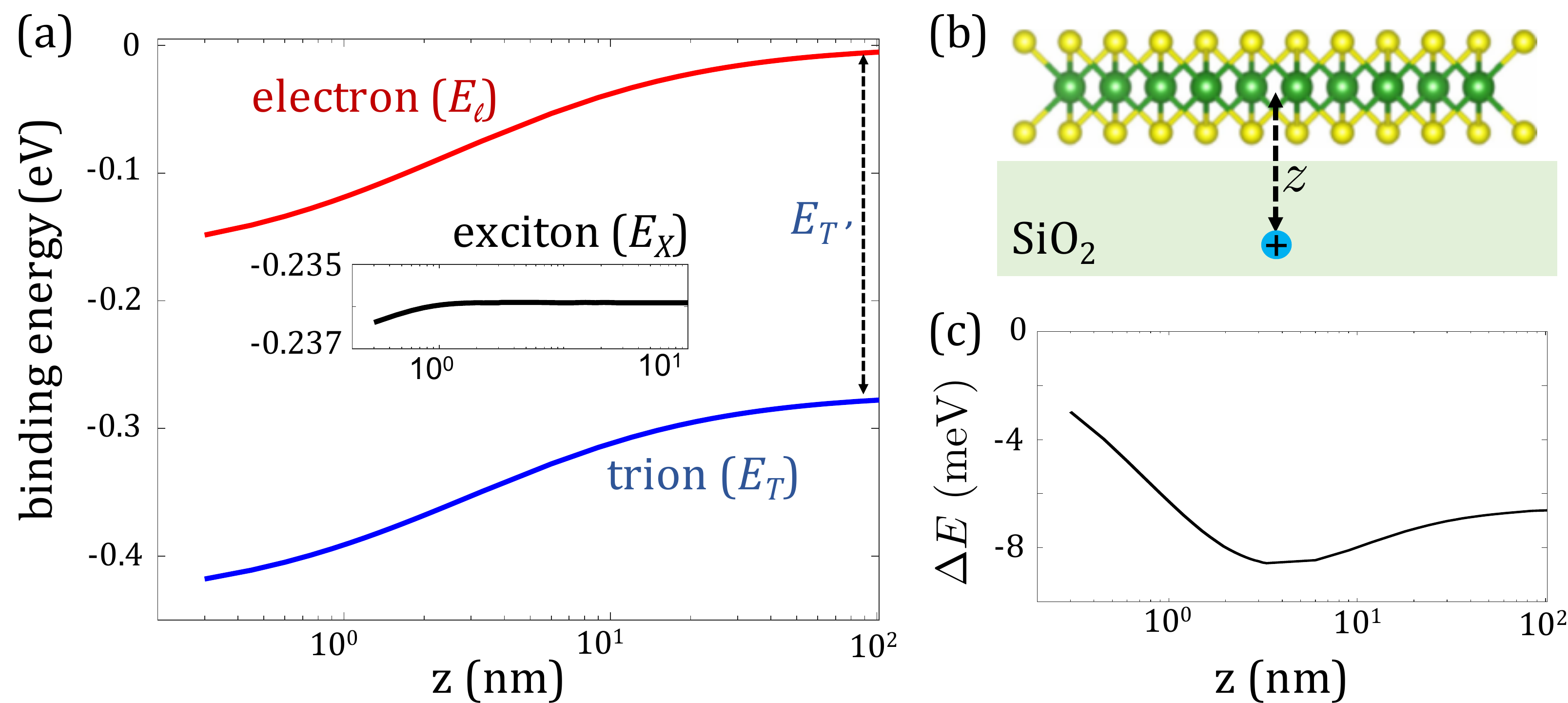}
 \caption{(a) Ground-state energies of electrons ($E_{\ell}$), excitons ($E_X$, inset), and negative trions ($E_T$) in ML-WSe$_2$ supported on SiO$_2$ substrate as a function of the distance of a positive point-charge defect  from the mid-plane of the ML. $E_{T'}$ denotes the energy of delocalized trions ($z \rightarrow \infty$). (b) A cartoon of the ML, substrate and extrinsic defect. (c) $\Delta E =(E_T+E_{\lambda}) - (E_X+E_{\ell})$, where $E_{\lambda}$$\,$$=$$\,$30~meV is the phonon energy. } \label{fig:binding}
\end{figure}

We begin the analysis by solving the Schr\"{o}dinger Equation of an $\mathcal{N}$-particle system with the  Hamiltonian
\begin{eqnarray}
H_{\mathcal{N}} =  \sum_i^{\mathcal{N}} V_{\text{E}}(r_{i},z) - \frac{\hbar^2}{2m_i}\nabla^2_i  + \sum_{i<j}^{\mathcal{N}} V_{\text{I}}(r_{ij})  \,.  \label{eq:H}
\end{eqnarray}
The particles are influenced by a point-charge defect in the substrate whose distance from the mid-plane of the ML  is $z$, as shown by Fig.~\ref{fig:binding}(b). $V_{\text{E}}(r_{i},z)$ is the Coulomb interaction between this extrinsic defect and the $i^{\text{th}}$ particle in the ML. The effective mass of the latter  is $m_i$. $V_{\text{I}}(r_{ij})$ is the Coulomb interaction between the $i^{\text{th}}$ and $j^{\text{th}}$ particles where $r_{ij}=|{\bf r}_i -{\bf r}_j|$. It is relevant when $\mathcal{N} \geq 2$. The supplemental information includes details on the Coulomb interactions and parameter values we use in the simulations. Equation~(\ref{eq:H}) is solved for localized electrons ($\mathcal{N}$$=$1), excitons ($\mathcal{N}$$=$2), and trions ($\mathcal{N}$$=$3) by the Stochastic Variational Method (SVM)\cite{VanTuan_arXiv18,Varga_PRC95,Varga_CPC08,Mitroy_RMP13,Kidd_PRB16,Donck_PRB17}.  

Figure~\ref{fig:binding}(a) shows the calculated energies in ML-WSe$_2$ when a $+1e$ charged defect is embedded in SiO$_2$.  The electron binding energy is largest when the defect is at the surface, $E_{\ell}(z\!=\!0.3$$\,$nm$)$$\,$$\sim$$\,$150~meV.  A similar localization-induced enhancement is seen in the energy of the trion ($E_T$). On the other hand, excitons are essentially unaffected by the defect, as shown by the inset. The weak dependence stems from the confluence of exciton neutrality and lack of contact between extrinsic defects and electrons or holes in the mid-plane of the ML. Similarly, $E_T-E_{\ell}$ is also largely independent of $z$ because the defect interacts weakly with the added electron-hole pair in the trion complex. Figure~\ref{fig:binding}(c) shows the value of $\Delta E =(E_T+E_{\lambda}) - (E_X+E_{\ell})$, where $E_{\lambda}$$\,$$=$$\,$30~meV is the energy of the homopolar phonon.

%In addition to being small, $\Delta E$ does not vary significantly between cases of nearby and remote defects. 
%vEssentially,  $\Delta E \sim E_{T'}+E_{\lambda} -E_X$ where v$E_{T'}$ is the energy of delocalized trions, as indicated by the arrow in Fig.~\ref{fig:binding}(a).

%represents the missing energy needed to form a real localized trion state when a localized electron captures the exciton following emission of the homopolar phonon. The fact that $\Delta E$ is not exactly zero but very close to it, as seen from Fig.~\ref{fig:binding}(c), indicates that virtual trion states play an important role in phonon-assisted optical transitions of neutral excitons (when the density of localized electrons is non-negligible). The fact that $\Delta E$ is not exactly zero also indicates that generation of real localized trion states should be assisted by additional agents such as acoustic phonons to satisfy complete energy conservation. 

Next, we analyze interactions that facilitate the capture process. When the magnitude of their matrix elements is comparable or larger than $|\Delta E|$, the phonon-assisted optical transitions display a strong resonance in the emission spectra. The matrix element reads  
\begin{eqnarray}
M_{\lambda}(\mathbf{K},\mathbf{q})=  \langle \Psi_T  | \, \Sigma_j \beta_j D_{j,\lambda}(\mathbf{q})e^{i\mathbf{q}\mathbf{r}_j}  | \Psi_X(\mathbf{K})\Psi_{\ell}   \rangle  \,\,,\,\, \label{eq:exciton_phonon}
\end{eqnarray}
where $\lambda=\{A_1',E_2'\}$ is the phonon mode (Fig.~\ref{fig:exp}). $\mathbf{q}$ and $\mathbf{K}$ are the phonon and exciton wavevectors, respectively. The localized electron (trion) state is denoted by $\Psi_{\ell}$ ($\Psi_{T}$). The exciton in the initial state is delocalized, $ \Psi_X(\mathbf{K}) =  \varphi_X(r) e^{i\mathbf{K}\cdot \mathbf{R}}/\sqrt{A}$, where $r=|\mathbf{r}_e-\mathbf{r}_h|$ and $\mathbf{R}=(m_e\mathbf{r}_e + m_h\mathbf{r}_h)/(m_e+m_h)$  are the relative and center-of-mass coordinates, respectively. $A$ is the area of ML. The sum over $j$ takes into account the long-range Fr\"{o}hlich interaction  between the $j^{\text{th}}$ particle and longitudinal-optical phonons, $D_{j,E_2'}(\mathbf{q})$, or the short-range interaction of the particle with homopolar phonons, $D_{j,A_1'}(\mathbf{q})$.  $\beta_j$$\,$$=$$\,$$1$~($\beta_j$$\,$$=$$\,$$-1$) when the $j^{\text{th}}$ particle is an electron (hole). The supplemental information file includes technical details of these interactions, along with a compiled list of all the parameters we use. % in Table S1 of the supplemental information file, along with results for the case that the ML is encapsulated in hBN. 

\begin{figure}[b]
\includegraphics[width=8.7cm]{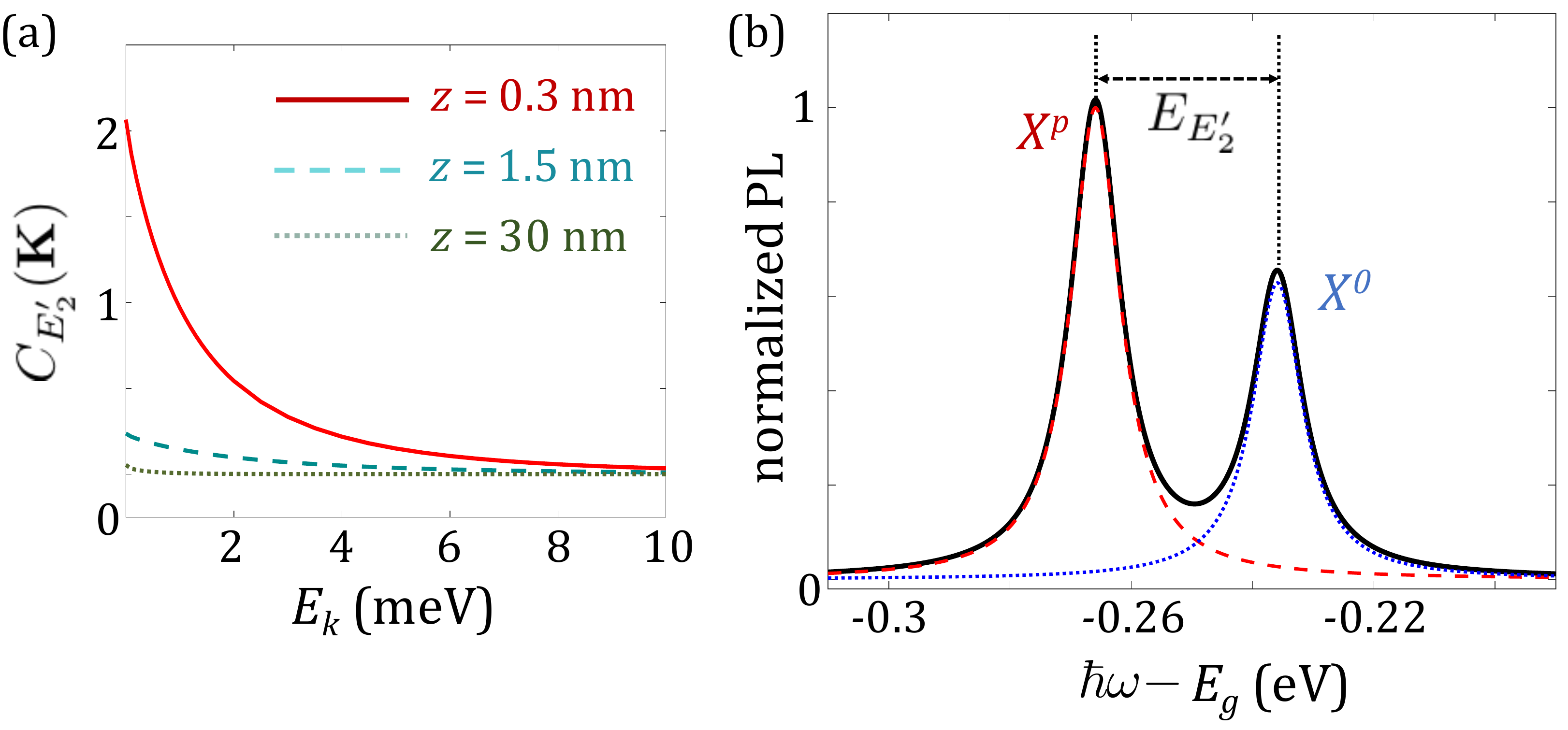}
 \caption{(a) The capture parameter, $C_{E_2'}(\mathbf{K})$, as a function of the exciton kinetic energy when $n_d=10^{12}$~cm$^{-2}$. (b) The normalized PL spectrum at $T$$\,$$=$$\,$5~K when $z$$\,$$=$$\,$0.3~nm and $\tau_{\ell} = 100\tau_{X}$. To account for disorder in the ML, an inhomogeneous broadening of 5~meV has been introduced. The simulations use the parameters of ML-WSe$_2$ supported on SiO$_2$ substrate. } \label{fig:MR}
\end{figure}

The rate of the phonon-assisted recombination process next to localization centers reads \cite{supp}
\begin{eqnarray}
\frac{1}{\tau_{\lambda}(\mathbf{K})} =   \left[ \frac{ n_dA \sum_{ \mathbf{q}}   \left| M_{\lambda}(\mathbf{K},\mathbf{q}) \right |^2  }{ (\Delta E - E_K )^2 } \right] \frac{ 1}{\tau_{\ell}} \,, \label{eq:tau_p}
\end{eqnarray}
where $n_d$ is the density of localized electrons due to extrinsic defects, and $E_K = \hbar^2K^2/2M$ is the exciton kinetic energy  ($M=m_e+m_h$). $\tau_{{\ell}}$ is the radiative decay time of a localized trion. The expression in square brackets, hereafter denoted by $C_{\lambda}(\mathbf{K})$, is unitless and it represents the amplitude of the virtual  capture process. %$\Gamma_{h}$ denotes homogenous broadening due to exciton scattering. 
Figure~\ref{fig:MR}(a) shows its value for the Fr\"{o}hlich interaction, $C_{E_2'}(\mathbf{K})$, when   $n_d=10^{12}$~cm$^{-2}$.  We see that faster excitons are less prone to the capture process, and that the amplitude is much weaker when the defect is remote. The latter reassures the important role of localization. It is so because $M_{\lambda}(\mathbf{K},\mathbf{q})$ is sizable when $q\ell_e \lesssim 1$, where $\ell_e$ is the characteristic electron localization length. That is, tighter localization enables more phonons to be involved in the capture process and contribute effectively to the sum in Eq.~(\ref{eq:tau_p}). In addition, the overlap between the initial and final states in Eq.~(\ref{eq:exciton_phonon}) is optimal for nearby defects because $\ell_e$ becomes comparable to both the exciton Bohr radius  and the characteristic distances between the particles of the trion complex.

%We have assumed that the phonon energy is $\mathbf{q}$-independent because of the relatively flat energy branches of optical phonons. 

%Note that $M^2_{\lambda}(\mathbf{k},\mathbf{q}) \propto 1/A$ because the electron and trion are localized whereas the exciton is not. On the other hand, $\widetilde{M}_{\lambda}^2(\mathbf{K})\propto n_d$.   

Note that while the amplitude of the capture process depends on how far the extrinsic defect is from the ML, as shown in Fig.~\ref{fig:MR}(a), the energy of the emitted photon does not: The spectral line always appears at $E_{\lambda}$ below $X_0$. %$\hbar \omega \simeq E_g - |E_X| - E_{\lambda}$, where $E_g$ is the band-gap energy. Recall that $\Delta E =(E_T+E_{\lambda}) - (E_X+E_{\ell})$ in the denominator of Eq.~(\ref{eq:tau_p})  has weak dependence on $z$ because $E_T-E_{\ell} \sim E_{T'}$ (see Fig.~\ref{fig:binding}). 
Consequently, the resonance photon energies are similar in devices in which the defects are concentrated at a certain distance from the ML, and others in which the defects are uniformly distributed in the substrate. In addition, one should note that the signature of intrinsic charged defects (inside the ML) should mostly be looked for in the low-energy side of the emission spectrum. The reason is that such defects strongly localize the excitons and increase their binding energies.

We compare the rates of phonon-assisted recombination next to localization centers and direct recombination of delocalized excitons that are not subjected to localization effects ($X^0$ in Fig.~\ref{fig:exp}). The effective rate of the phonon-assisted process follows from the average
\begin{eqnarray}
 \Bigg \langle \frac{1}{ \tau_{\lambda}(\mathbf{K})  } \Bigg\rangle=   \frac{\sum_{\mathbf{K}} f_X(\mathbf{K})/\tau_{\lambda}(\mathbf{K})}{\sum_{\mathbf{K}} f_X(\mathbf{K})} , \label{eq:avg}
\end{eqnarray}
where $f_X(\mathbf{k})$ is the exciton distribution function. The effective rate of the direct recombination reads \cite{Wang_PRB16}
\begin{eqnarray}
\frac{1}{\tau_0} =   \left[ \frac{ (\hbar\omega_0)^2 }{2 k_BT Mc^2 } \right] \frac{1}{\tau_X} \,. \label{eq:tau_o}
\end{eqnarray}
$\hbar\omega_0 = E_g - |E_X|$ is the resonance photon energy where $E_g$ is the band-gap energy. $c$ is the speed of light in vacuum and $k_BT$ is the thermal energy. $\tau_X$ is the radiative decay time of delocalized excitons in the light cone. It is typically the fastest decay process, where $\tau_X \ll \tau_{\ell}$ \cite{Wang_PRB16}. The term in square brackets denotes the strong attenuation after averaging over the distribution of excitons. It is about 0.01 already at $T=5$~K, reflecting the fact that only a negligible fraction of delocalized excitons reside in the light cone (excitons whose kinetic energy is of the order of a few $\mu$eV). We do not consider the acoustic-phonon-assisted recombination of delocalized excitons in our analysis (`pushing' the excitons into the light cone by emission of low-energy phonons). The main effect of acoustic phonons is to broaden the high-energy tail of $X^0$ and not to strongly amplify the emission \cite{Shree_arXiv18}.

Figure~\ref{fig:MR}(b) shows the calculated normalized PL for a Boltzmann distribution of excitons at $T$$\,$$=$$\,$5~K.  %The band-gap energy, $E_g$, is the reference level of the $x$-axis. 
The intensity of the phonon-related peak ($X^p$) is commensurate with $1/\tau_{\ell}$, and it depends on the defects' density and distance from the ML. The intensity of the delocalized exciton peak ($X^0$) is commensurate with $1/\tau_X$. We have used $n_d=10^{12}$~cm$^{-2}$, assumed surface defects ($z$$\,$$=$$\,$0.3~nm), and considered the Fr\"{o}hlich interaction ($\lambda=E_2'$). Figure~\ref{fig:MR}(b)  shows that the emission from delocalized excitons in the light cone is weaker even when assuming that their radiative decay is 100 times faster than that of localized trions, $\tau_{\ell} = 100\tau_{X}$. This result provides a strong evidence for the importance of phonon-assisted recombination next to localization centers in ML-TMDs.

% Comparing Fig.~\ref{fig:MR}(b) with  the experimental data in Fig.~\ref{fig:exp}, we find a noticeable difference in the linewidth of the peaks. The reason is that we do not include inhomogeneous broadening due to disorder. In this view, the results of  Fig.~\ref{fig:MR}(b) better reflect the spectra from relatively clean samples \cite{Courtade_PRB17,Ajayi_2DMater17}.   

% ($\nu_{\text{LA}}=5\cdot10^2$~cm/s and $\nu_{\text{TA}}=5\cdot10^2$~cm/s in WSe$_2$ \cite{sound velocity}). 
% d=0.60 \ell_+ / \ell_- = 0.8;  alpha=1.9 where  2*\ell_-   +   \ell_+= alpha*  epsilon_ML*d/2

Before we discuss the implications of the results, it is emphasized that optical transitions next to localization centers do not play a significant role in absorption-type experiments such as differential reflectance spectroscopy. In these measurements, one probes the oscillator strength of delocalized exciton and trion states. The signature of defects in the absorption spectrum is manifested through broadening effects. %, while other effects are not at play as long as the density of impurities is not exceedingly large (i.e., as long as the average distance between impurities remains much larger than the exciton Bohr radius).
On the other hand, radiative recombination next to localization centers is significant in the emission process because thermalized excitons can be captured before entering the minuscule light cone region. 

We now turn to discussion of the results, focusing first on the distinction between radiative recombinations that involve real and virtual trions. The latter requires emission of an optical phonon during the radiative process, rendering the distinction straightforward if the energy difference between $E_X-E_{\lambda}$ and $E_T-E_{\ell}$ is noticeable. This scenario is somewhat applicable in sulfur-based MLs. For example, recombination with virtual trions  in ML-WS$_2$ has a spectral line at $E_{\lambda}$$\,$$\sim$$\,$44~meV (356~cm$^{-1}$) below $X^0$, whereas the spectral line from real trions has a fine-structure doublet at $\sim$30 and  $\sim$37~meV below $X^0$ \cite{Plechinger_NatCommun16}. The distinction is more subtle in ML-WSe$_2$ where the phonon-related spectral line nearly resonates with one of the doublet features ($\sim$$\,$30~meV below $X^0$) \cite{Jones_NatPhys16,Courtade_PRB17}. In this case, the gate-voltage dependence of the PL can be used to identify the recombination process. The phonon-related peak is often observed at small gate voltages whereas the trion doublet emerges at large gate voltages \cite{Courtade_PRB17,Zhou_NatNano17,Wang_RMP18}. The reason is that increasing the electron density by the gate eventually screens the charged defects and electrons become delocalized. The capture process is then suppressed. In this limit, the doublet feature in the PL comes from delocalized trions. 

The distinction between radiative recombinations with real and virtual trions is more subtle in MoSe$_2$ where the emission from trions has a single feature. However, we can still recognize the effect of localization by comparing the absorption and emission spectra of a  gated device. If the suspected peak appears in both spectra at the same voltage level, then it should be associated with delocalized trion states. On the other hand, it should be associated with localized trion states if it is absent in the absorption spectrum.   Clearly, removing extrinsic defects from the substrate suppresses the recombination next to localization centers. This behavior was indeed observed by Ajayi \textit{et al.}, who showed that treatment of the SiO$_2$ substrate led to substantial suppression of the low-energy peak in the emission spectrum \cite{Ajayi_2DMater17}. 

Other than the dependencies on sample quality and gate voltage, localization effects can be inferred from the temperature dependence of the PL. The recombination next to localization centers decays with temperature because (i) excitons gain kinetic energy making them less susceptible to the capture process, as shown in Fig.~\ref{fig:MR}(a), and (ii) electrons escape the localization center. The decay of delocalized trions with temperature, on the other hand, is governed by thermal dissociation of the complex into delocalized electron and exciton. PL experiments in ML-TMDs often show that the decay already occurs when $k_BT \ll (E_X - E_{T'})$, where $E_{T'}$ is the energy of a delocalized trion. This behavior implies a dominant contribution from recombination next to localization centers, as was recently suggested by Godde  \textit{el al.} \cite{Godde_PRB16}. In addition, it is supported by the fact that $\mu$-PL experiments show that the emission of the low-energy peak is localized in nature compared to the spatially homogeneous emission of neutral excitons \cite{Bao_NatComm15,private}. 
 %These findings reinforce our proposed phonon-assisted mechanism. The temperature decay in this case is governed by thermal escape of the electrons from their localization centers and by the kinetic energy of the neutral exciton (faster excitons are less susceptible to the capture process, as shown in Fig.~\ref{fig:Mk2}). As a result, the decay of the phonon-assisted process is governed by the binding energy of the electrons to the defects and by the thermal kinetic energy of neutral excitons.

Applying a strong magnetic field is additional way to identify the localization effect. Plechinger \textit{et al.} have recently demonstrated a strong amplification of the phonon-assisted optical transition in ML-WS$_2$ supported on SiO$_2$ when the applied  out-of-plane magnetic field was about $30$~T \cite{Plechinger_NanoLett16}. These findings reinforce our proposed mechanism because of the equivalent effects of a strong magnetic field and closer impurity. Namely, the localization of electrons is enhanced due to a smaller cyclotron radius, or equivalently, if the impurity is closer to the ML.  Both effects render the capture process more effective and the phonon-assisted  recombination stronger.  
 
In conclusion, we have unravelled an intriguing  phonon-assisted radiative process in the photoluminescence of monolayer transition-metal dichalcogenides. It starts when a localized electron virtually captures the exciton by emitting a phonon, followed by photon emission from the intermediate  virtual trion state. Overall, it is a strong process because $E_T - E_{\ell}$ nearly resonates with $E_X-E_{\lambda}$ where $E_{T(\ell)}$ is the energy of a localized trion (electron) and $E_{X(\lambda)}$ is the energy of an exciton (optical phonon). It is this resonance condition that led to widespread confusion, where the radiative recombination was often attributed to real trions instead of virtual ones; participation  of optical phonons in the radiative process was largely ignored. We have discussed ways to distinguish between radiative recombinations with real and virtual trions based on the energy of the emitted photons. Furthermore, ways to probe localization effects were analyzed by comparing the absorption and emission spectra, as well as by using device quality, gate voltage, temperature, and magnetic field as knobs in photoluminescence experiments. %, or the comparison with absorption spectrum 

\acknowledgments{The work at the University of Rochester was mostly supported by the Department of Energy, Basic Energy Sciences, under Contract No. DE-SC0014349, whereas the computational work was also supported by the National Science Foundation (Grant No. DMR-1503601). Work at the University of Washington was  supported by the Department of Energy, Basic Energy Sciences, Materials Sciences and Engineering Division (DE-SC0018171).}

\begin{widetext}

\section{Supplemental information: Virtual trions in the photoluminescence of monolayer transition-metal dichalcogenides}

%  The phonon energy is close to but not necessarily the exact one needed for transition into the intermediate virtual state, and this resemblance often leads to widespread confusion where the optical transition is attributed to a real trion instead of a phonon-assisted process. We analyze the differences between the two mechanisms, providing useful ways to identify the peaks in the emission spectrum.

This supplemental information file includes
\begin{enumerate}
\vspace{-1mm}
\item Details of the Coulomb potentials we use in Eq.~(1) of the main text. We analyze the Coulomb interaction between the electron (or hole) in the ML and the extrinsic defect, as well as the Coulomb interaction between charged particles in the ML. 
\vspace{-1mm}
\item Details of the exciton-phonon matrix elements in Eq.~(2) of the main text. It includes the Fr\"{o}hlich interaction and short-range interaction with optical phonons (thickness fluctuations). 
\vspace{-1mm}
\item Derivation of Eq.~(3) in the main text.   
\vspace{-1mm}
\item A compiled list of all the parameters we use in the simulations.
\vspace{-1mm}
\end{enumerate}

\section{Coulomb interactions ($V_{\text{I}}$ \& $V_{\text{E}}$ in Eq.~(1) of the main text)}
Empirically, the optical phonon energies in ML-TMDs nearly resonate with the energy difference between the spectral lines of delocalized trions and excitons. This resonance is crucial for the explanation of the phonon-assisted recombination next to localization centers.  Therefore, the most important requisite is that the Coulomb potential between charged particles in the ML, $V_{\text{I}}$, can recover this resonance when we solve the Schrodinger Equation for delocalized excitons and trions. 

As we have recently shown, the widely used Rytova-Keldysh potential has problems in modeling trion states \cite{VanTuan_arXiv18}. For example, experimental results show that the energy difference between the spectral lines of delocalized trions and excitons is hardly changed when the ML is suspended in air, supported on SiO$_2$, or encapsulated in hBN. The Keldysh potential cannot recover this behavior. To solve this problem we have derived an alternative potential form that considers the polarizability in each of the three atomic sheets that compose the ML, as shown in Fig.~\ref{fig:3chi}(b).  Employing the new potential, $V_{\text{I}}=V_{3\chi}$, we were able to achieve very good agreement with experiment in terms of trion and exciton binding energies in various environments \cite{VanTuan_arXiv18}.

\subsection*{Coulomb interaction between particles in the mid-pane of the ML, $V_{\text{I}}$}

\begin{figure}
\includegraphics[width=15.0cm]{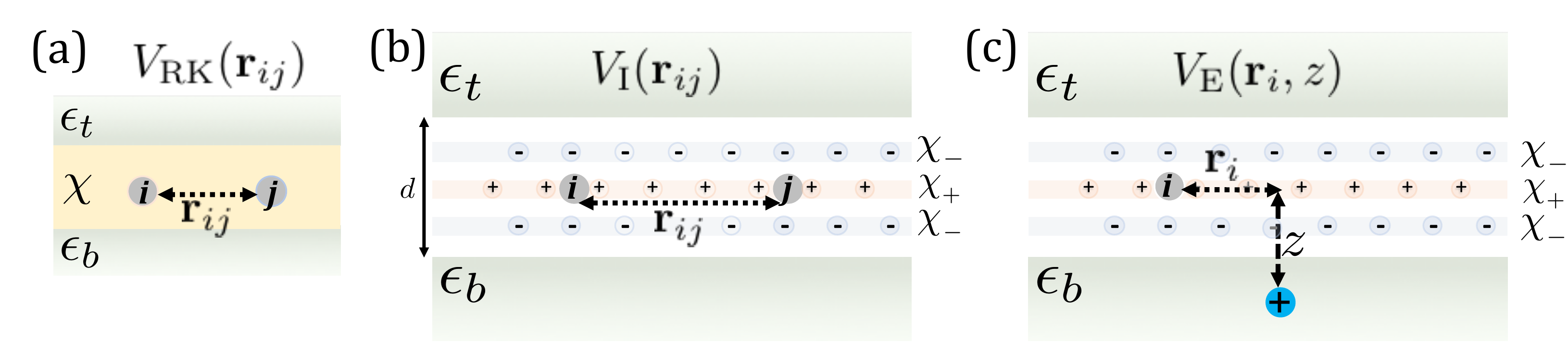}
 \caption{(a)  The dielectric environment when considering a uniform ML with polarizability $\chi$. The solution of the Poisson Equation in the ideal 2D limit (zero thickness of the ML) leads to the celebrated Rytova-Keldysh potential, $V_{\text{RK}}(\mathbf{r}_{ij})$ between the charges $i$ and $j$. (b) The revised ML geometry when modeling the Coulomb interaction between the charged particles in the mid-plane of a ML of thickness $d$. The ML is viewed as three atomic sheets with polarizabilities $\chi_+$ for the central one (Mo/W) and $\chi_-$ for the top and bottom ones (S/Se/Te, displaced by $\pm$d/4 from the center).  Screening from the chalcogen  sheets helps to confine the field lines in the ML, thereby reducing the dependence on the bottom and top materials whose dielectric constants are $\epsilon_b\,\&\,\epsilon_t$. (c) The same as in (b) but when modeling the Coulomb interaction between a charged particle in the mid-plane of a ML and an extrinsic positive-charged defect embedded in the bottom dielectric layer. }
  \label{fig:3chi}
\end{figure}

Regardless of the model we use for our 2D system, the solution of the Poisson Equation leads to the following 2D Fourier transform of the Coulomb potential,
\begin{equation}
 V_I({ q})=\frac{2\pi e_ie_j}{A\epsilon_\text{I}(q)q}\,\,, \label{eq:2D_potential_Fourier}
 \end{equation}
where $e_i$ and $e_j$ are the charges of the interacting particles, and $A$ is the area of the ML. The static dielectric function, $\epsilon_\text{I}({q})$, depends on the model we use. In the Rytova-Keldysh case \cite{Cudazzo_PRB11}
\begin{equation}
 \epsilon_\text{I}({\bf q}) =   \epsilon_\text{RK}({\bf q}) = \frac{\epsilon_b + \epsilon_t}{2} + r_0 q, \label{Eps_RK}
 \end{equation}
where $\epsilon_{b(t)}$ is the dielectric constant of the bottom (top) dielectric layer, and $r_0=2\pi\chi$ is the screening length in the ML due to its polarizability, $\chi$. If we use the  geometry shown in Fig.~\ref{fig:3chi}(b), we get \cite{VanTuan_arXiv18}
\begin{equation}\label{EpsI}
\epsilon_{\text{I}}(q) =   \epsilon_{3\chi}({\bf q}) =\frac{1}{2}\left[\frac{N_t(q)}{D_t(q)}+\frac{N_b(q)}{D_b(q)}\right],  
\end{equation}
where
\begin{eqnarray}
D_\gamma(q) &=& 1+q\ell_- -q\ell_- (1+p_\gamma)\text{e}^{-\frac{qd}{2}} - (1-q \ell_- ) p_\gamma \text{e}^{-qd}, \nonumber \\
N_\gamma(q) &=& \left(1+q\ell_-\right)\left(1+q\ell_+\right) \nonumber  +  \left[\left(1-p_\gamma\right)-\left(1+p_\gamma\right)q\ell_+\right]q\ell_-\text{e}^{-\frac{qd}{2}}  + (1-q\ell_-)(1-q\ell_+ )p_\gamma\text{e}^{-qd}.
\end{eqnarray}
$d$ is the thickness of the ML, $p_\gamma \equiv (\epsilon_\gamma-1)/(\epsilon_\gamma+1)$ for $\gamma=\{b,t\}$, and $ \ell_\pm=2\pi\chi_\pm$ are the dielectric screening lengths of each of the atomic sheets in the ML, as shown in  Fig.~\ref{fig:3chi}(b). Note that $\epsilon_{3\chi}({\bf q}) = \epsilon_{\text{RK}}({\bf q}) $ when $d=0$ and $r_0 = \ell_+ + 2\ell_-$. Figure~\ref{fig:eps_pot}(a) shows the inverse dielectric functions  for a ML supported on SiO$_2$ and exposed to air, $\epsilon_b=3.9$ and $\epsilon_t=1$. The solid (dashed) line shows the 3$\chi$ (Rytova-Keldysh) model. Their values converge to $2/(\epsilon_b+\epsilon_t)$ when $q \rightarrow 0$. 

\begin{figure}
\includegraphics[width=15.0cm]{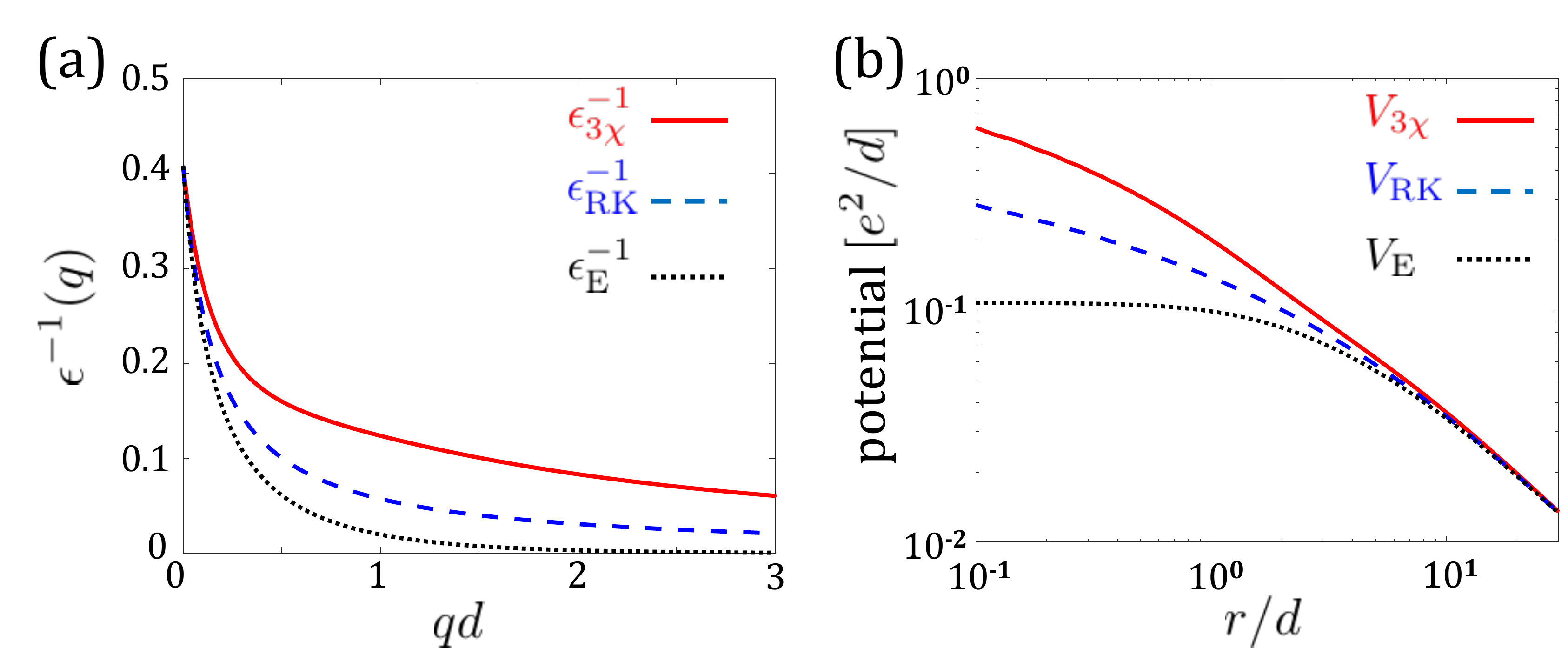}
 \caption{(a)  Inverse of the static dielectric functions, following Eqs. (\ref{Eps_RK}), (\ref{EpsI}), and (\ref{Eq:BareCoulomb}). (b) The resulting potential forms in real space. The parameters we use are $d=0.6$~nm, $\epsilon_t=1$ (air), $\epsilon_b=3.9$ (SiO$_2$), $\ell_+=\ell_-=5d$, $r_0=15d$. For the extrinsic impurity case (dotted black lines), we assume a defect at the SiO$_2$ surface, $z=0.5d$. } \label{fig:eps_pot}
\end{figure}

The real-space Coulomb potential then follows from
\begin{equation}
V_{\text{I}}({r}) =  e_ie_j \int_0^{\infty}  \! \! dq \frac{ J_0(qr)}{\epsilon_{\text{I}}(q)} \,\, , \,\,\,\, \label{eq:2D_real}
 \end{equation}
where $J_0$ is the zeroth-order Bessel function. Figure~\ref{fig:eps_pot}(b) shows the real-space potential forms using the dielectric static functions in Figure~\ref{fig:eps_pot}(a). $V_{\text{RK}}$ and $V_{3\chi}$ converge to the conventional potential $2e_ie_j/(\epsilon_b+\epsilon_t)r$ when $r > \{d,r_0,\ell_{\pm}\}$. Their values deviate at short distances, $r \lesssim d$, corresponding to the region $1  \lesssim qd$ in  Fig.~\ref{fig:eps_pot}(a).

\subsection*{Coulomb interaction between a particle in the mid-pane of the ML and an extrinsic defect, $V_{\text{E}}$}

Repeating the steps used to derive $V_{3\chi}$, we can similarly derive the Coulomb potential between a charged particle in the mid-plane of the ML ($z'=0$) and a charged extrinsic defect embedded in the bottom (or top) dielectric layers.  Its solution comes from the relation, $V(q,z)= e_i \phi_{\bm{q}}(z'=0,z)$, where $\phi_{\bm{q}}(z'=0,z)$ is the 2D Fourier transform of the extrinsic-defect-induced potential. The Poisson Equation in this case reads
\begin{eqnarray}
\frac{\partial}{\partial z'}\!\left[\!\kappa(z')\frac{\partial\phi_{\bm{q}}(z',z)}{\partial z'}\!\right]\!-\kappa(z')q^2\phi_{\bm{q}}(z',z)=-\frac{4\pi e_d}{A}\delta\left(\!z'\!-\!z \right) +2 q^2 \! \left[\delta(z')\ell_+ + \delta\!\left(\!z'\!-\!\tfrac{d}{4}\right)\!\ell_- + \delta\!\left(\!z'\!+\!\tfrac{d}{4}\right)\!\ell_- \! \right]\!\phi_{\bm{q}}(z',z), \label{Eq:Poisson_Ext}
\end{eqnarray}
where $e_d$ denotes the charge of the defect, $z$ its distance from the mid-plane of the ML, and
\begin{equation}\label{Eq:DielEnv2}
\kappa(z')=\left\{\begin{array}{ll}
 \epsilon_t & \mathrm{for}\quad z'>d/2,\\
 1 & \mathrm{for}\quad -d/2<z'<d/2,\\
 \epsilon_b & \mathrm{for}\quad z'<-d/2\,.
 \end{array}\right.
\end{equation}
Fixing the point-charge defect to the bottom layer, $z < -d/2$,  one can solve Eq.~(\ref{Eq:Poisson_Ext}) with the boundary conditions that $\phi_{\bm{q}}(z',z)$ is continuous and its derivative is piecewise continuous with jumps of $2q^2\ell_+\phi_{\bm{q}}(0,z)$ at $z'=0$, $2q^2\ell_- \phi_{\bm{q}}(\pm d/4,x)$ at $z'=\pm d/4$, and $-4\pi e_d/A$ at $z'=z$. The Coulomb interaction between $e_d$ and a charged-particle $e_i$ in the mid-plane of the ML yields  
\begin{equation}
V(q,z)= e_i \phi_{\bm{q}}(z'=0,z) = \frac{2\pi e_de_i}{A\epsilon_{\text{E}}(q,z)q}\,\,,\qquad \,\,\,  \epsilon_{\text{E}}^{-1}({\bf q},z)=\frac{2\left(c_0 + c_1 e^{\frac{qd}{2}} + c_2 e^{qd}\right)  e^{-q(z-d)} }{ d_0 + d_1 e^{\frac{qd}{2}} + d_2 e^{qd}+ d_3 e^{\frac{3qd}{2}} + d_4 e^{2qd}} \,,  \label{Eq:BareCoulomb} \,\,\,
\end{equation}
where 
\begin{eqnarray}
c_0&=& \left(q\ell_- -1\right) \left(\epsilon_b-1\right),  \nonumber  \\
c_1&=& -2 q\ell_- \epsilon_b , \nonumber \\
c_2&=&  \left(q\ell_- +1\right) \left(\epsilon_b+1\right),
\end{eqnarray}
and
\begin{eqnarray}
d_0&=&  \left(q\ell_- -1\right){}^2 \left(q\ell_+-1\right) \left(\epsilon_t-1\right) \left(\epsilon_b-1\right) ,\nonumber \\
d_1&=& -2q\ell_- \left(q\ell_- -1\right)  \left[1-q\ell_+(\epsilon_t+ \epsilon_b) +  \left(2q\ell_+ -1\right) \epsilon_t\epsilon _b \right],  \nonumber\\
d_2&=&  -2 q^2\ell_-^2 \left(\epsilon_t+\epsilon_b\right)+2 q\ell_+ \left[1-q^2\ell_-^2+\left(3 q^2\ell_-^2 -1 \right) \epsilon_t \epsilon_b\right], \nonumber  \\
d_3 &=& -2 q\ell_- \left(q\ell_- +1\right) \left[-1+q\ell_+(\epsilon_t+ \epsilon_b) + \left(2q\ell_+ +1\right) \epsilon_t\epsilon_b  \right], \nonumber \\
d_4&=& \left(q\ell_- +1\right){}^2 \left(q\ell_+ +1\right) \left(\epsilon_t+1\right) \left(\epsilon_b+1\right).
\end{eqnarray}
The dotted-black line in Fig.~\ref{fig:eps_pot}(a) shows the resulting inverse dielectric function for a surface defect, $z=d/2$, with charge $e_d=e$. All other values are the same as before: $d=0.6~nm$, $\ell_{\pm}=5d$, $\epsilon_t=1$ and $\epsilon_b=3.9$. The ensuing real-space 2D interaction,  
\begin{eqnarray}
V_{\text{E}}(r,z) =  e_de_i \int_0^{\infty}  \! \! dq \frac{ J_0(qr)}{\epsilon_{\text{E}}(q,z)} \,\, , \,\,\,\, \label{eq:2D_potential_Fourier}
\end{eqnarray}
is shown by the dotted-black line in Fig.~\ref{fig:eps_pot}(b). The potential converges to the conventional one $2e_de_i/(\epsilon_b+\epsilon_t)r$ when $r > \{d,z,\ell_{\pm}\}$. Its value saturates at short distances, $r < z$, because the potential scales as $1/\sqrt{z^2+r^2}$.  

%%%%%%%%%%%%%%%%%%%%%%%%%%%%%%%%%%%%%%%%%%%%%%%%%%%%%%%%%%%%%%%%%%%%%%%%%%%%%
%%%%%%%%%%%%%%%%%%%%%%%%%%%%%%%%%%%%%%%%%%%%%%%%%%%%%%%%%%%%%%%%%%%%%%%%%%%%%
%%%%%%%%%%%%%%%%%%%%%%%%%%%%%%%%%%%%%%%%%%%%%%%%%%%%%%%%%%%%%%%%%%%%%%%%%%%%%

\subsection*{Details of the simulation of Fig. 2}
We use the Stochastic Variational Method (SVM) to solve  Eq.~(1) of the main text \cite{Varga_PRC95,Varga_CPC08}. This method has been recently applied to study binding energies of exciton complexes in ML-TMDs \cite{VanTuan_arXiv18,Mitroy_RMP13,Kidd_PRB16,Donck_PRB17}. Another common method is the Quantum Monte Carlo \cite{Mostaani_PRB17,Foulkes_RMP01}, and the two methods  are similar in the sense that both use trial functions to minimize the ground state energy. The wavefunction in the SVM is expanded in a variational basis which includes correlated Gaussian functions. This variational basis is optimized in a random trial procedure to minimize the ground state energy of a  few-body systems. The Supplemental information file of Ref.~\cite{VanTuan_arXiv18} includes all of the necessary details needed to implement the method in the absence of extrinsic defect (i.e., only with $V_{\text{I}}=V_{3\chi}$). Adding the potential of the defect, $V_{\text{E}}$, is straightforward \cite{Varga_CPC08}. The parameter values we use for  $m_e$, $m_h$, $d$, $\epsilon_{b}$, $\epsilon_{t}$, and $\ell_{\pm}$ in the simulations of Fig.~2 of the main text are listed in Sec.~\ref{sec:parameters}.  We find that it is sufficient to use a few tens of correlated Gaussians to accurately describe the electrons and exciton states, and a few hundreds for the trion case. 

%%%%%%%%%%%%%%%%%%%%%%%%%%%%%%%%%%%%%%%%%%%%%%%%%%%%%%%%%%%%%%%%%%%%%%%%%%%%%
%%%%%%%%%%%%%%%%%%%%%%%%%%%%%%%%%%%%%%%%%%%%%%%%%%%%%%%%%%%%%%%%%%%%%%%%%%%%%
%%%%%%%%%%%%%%%%%%%%%%%%%%%%%%%%%%%%%%%%%%%%%%%%%%%%%%%%%%%%%%%%%%%%%%%%%%%%%

\section{exciton-phonon interactions in ML-TMDs}

Of the nine phonon modes in ML-TMDs, six belong to the optical branches. Two of which are strongly coupled to spin-conserving scattering of electrons or holes \cite{Kaasbjerg_PRB12, Song_PRL13}: The longitudinal  optical (LO) and out-of-plane transverse optical (ZO) phonons. The LO mode is denoted by  $E_2'$ and the ZO by $A_1'$. The insets in the lower panels of Fig.~1 of the main text show their corresponding atomic displacements. Below we describe the electron (or hole) interaction with these phonon modes. 

The matrix element in Eq.~(2) of the main text describes the formation of a localized trion when a localized electron captures a delocalized exciton by emission of an optical phonon.  It contains the interaction terms $D_{j,\lambda}(\mathbf{q})$ where $\mathbf{q}$ is the phonon wavevector, $j=e(h)$ represents the electron (hole) component of the exciton, and $\lambda =\{ E_2', A_1'\}$. The coupling to the $E_2'$ mode is governed by the Fr\"{o}hlich interaction \cite{Sohier_PRB16}
\begin{equation}
D_{E_2'}(q)= D_{e,E_2'}(q)= D_{h,E_2'}(q) =  \sqrt{n_{E_2'} + \frac{1}{2} \pm \frac{1}{2} } \sqrt{\frac{A_u}{A}}\sqrt{\frac{\hbar^2}{2M_xE_{E_2'}}}  \left( 1 + \sqrt{\frac{M_x}{M_m} }\right) \frac{2\pi Z_{E_2'} e^2}{A_u\epsilon_{\text{I}}(q)}   
\end{equation}
where $n_{E_2'} = 1/[\text{exp}(E_{E_2'}/k_BT)-1]$ is the Bose Einstein distribution. $E_{E_2'}$ is the phonon energy where we have neglected its weak dependence on $\mathbf{q}$ due to the dispersionless nature of long-wavelength optical phonons. The $\pm$ denotes the case of phonon emission (plus) or absorption (minus). $A$ and $A_u$ are the areas of the ML and unit cell, respectively.  $M_x$ and $M_m$ are the masses of the chalcogen and transition-metal atoms, respectively. $Z_{E_2'}$ is the Born effective charge describing the linear relation between the force on the atom and the macroscopic electric field. Conservation of charge implies that $Z_{E_2'}=Z_{m}=2Z_x$. $\epsilon_{\text{I}}(q)$ is the static dielectric function, and it can be described either by Eq.~(\ref{Eps_RK}) or (\ref{EpsI}) because both forms are nearly identical in the long wavelength regime (where the interaction is strong), as shown by the solid and dashed lines in Fig.~\ref{fig:eps_pot}(a).  

The coupling of electrons and holes to ZO phonons ($A_1'$ mode) is governed by the short-range potential induced by the volume change of the unit-cell volume. This coupling can be viewed as the scattering that electrons or holes experience due to thickness fluctuations of the ML in the long-wavelength limit.  The corresponding interaction terms read \cite{Song_PRL13,Sohier_PRB16}
\begin{equation}
D_{j,A_1'}(q) \simeq D_{j,A_1'} = \sqrt{n_{A_1'} + \frac{1}{2} \pm \frac{1}{2} } \sqrt{ \frac{\hbar^2 A_u}{2A(2M_x+M_m) E_{A_1'} }}  \mathcal{S}^{(A_1')}_{j} 
\end{equation}
where the Bose Einstein distribution in this case is, $n_{A_1'} = 1/[\text{exp}(E_{A_1'}/k_BT)-1]$, and as before, we have neglected the weak $\mathbf{q}$-dependence of the phonon energy ($E_{A_1'}$) due to the dispersionless nature of long-wavelength optical phonons. $\mathcal{S}^{(A_1')}_{j}$ is the scattering constant of electrons ($j=e$) or holes ($j=h$).  Its amplitude is of the order of 1~eV/$\AA$.

\subsection*{Details of the simulation of Fig. 3}
The simulations in Fig.~3 of the main text are performed by using the following assumptions.
\begin{enumerate}
\item We have only considered the Fr\"{o}hlich interaction for the optical case because of its ultra-strong effect in ML-TMDs \cite{Sohier_PRB16}. Adding the contribution of the $A_1'$ mode should not lead to qualitative changes in the calculated PL because we simulate the case of ML-WSe$_2$ in Figs.~2 and 3 of the manuscript, for which the energies of the $A_1'$ and $E_2'$ modes nearly resonate (as shown by the Raman spectra in Fig.~1 of the manuscript). 

\item We have only considered spontaneous phonon emission,
\begin{eqnarray}
\sqrt{ \frac{1}{\text{exp}(E_{E_2'}/k_BT)-1} + \frac{1}{2} \pm \frac{1}{2}} &\rightarrow& 1. \nonumber
\end{eqnarray}
Given that $E_{E_2'}$ is of the order of a few tens of meV in all ML-TMDs, this approximation is readily justified at low temperatures. 
\end{enumerate}
The resulting matrix element that we have used in the simulations of Fig.~3 of the main text reads
\begin{eqnarray}
\!\!\!\!\!\!\!\!\!\!\!\! M_{E_2'}(\mathbf{K},\mathbf{q}) &\!\! = \!\!& \sqrt{\frac{A_u}{A}}\sqrt{\frac{\hbar^2}{2M_xE_{E_2'}}}  \! \left( \!1 \!+\! \sqrt{\frac{M_x}{M_m} }\right) \! \frac{2\pi Z_{E_2'} e^2}{A_u\epsilon_{3\chi}(q)}  \left\langle \!\! \Psi_T(\mathbf{r_h}, \mathbf{r_e}, \mathbf{r_{\ell}})  \left|  e^{i\mathbf{q}\mathbf{r}_{\ell}} \!+\! e^{i\mathbf{q}\mathbf{r}_{e}} \!-\! e^{i\mathbf{q}\mathbf{r}_{h}}   \right| \Psi_X(\mathbf{r_h}, \mathbf{r_e}; \mathbf{K}) \Psi_{\ell}(r_{\ell})  \!\! \right\rangle  \!,   \label{eq:main_text_2}
\end{eqnarray} 
where
\begin{eqnarray} 
\Psi_X(\mathbf{r_h}, \mathbf{r_e}; \mathbf{K}) &=& \frac{\exp({i\mathbf{K}{\mathbf{R}}})}{\sqrt{A}}\varphi(r) \,\,, \qquad r=|\mathbf{r_e}-\mathbf{r_h}| \,\,, \qquad \qquad \qquad  \mathbf{R}=  \alpha_e \mathbf{r}_e + \alpha_h \mathbf{r}_h \,\, .\nonumber
\end{eqnarray} 
We have used the SVM to calculate the localized electron and trion states, $\Psi_{\ell}(r_{\ell})$ and $\Psi_T(\mathbf{r_h}, \mathbf{r_e}, \mathbf{r_{\ell}})$, as well as the delocalized exciton state, $\varphi(r)$. These wavefunctions are expressed as sums of correlated Gaussians \cite{Varga_PRC95,Mitroy_RMP13}, 
\begin{eqnarray} 
\Psi_N(\mathbf{r}_1, \mathbf{r}_2, ...., \mathbf{r}_N) &=& \sum_i^n C_i \exp\left( -\frac{1}{2} \sum_{k<l}^N \alpha_{kl}^i r_{kl}^2 -\frac{1}{2} \sum_{k}^N \beta_{k}^i r_k^2   \right) \label{eq:svm_form}
\end{eqnarray} 
where $N$ is the number of particles in the complex, $n$ is the basis size, and $r_{kl} = |\mathbf{r}_k - \mathbf{r}_l |$. The variational parameters found by the SVM are $C_i$, $\alpha_{kl}^i$ and $\beta_{k}^i$. A great advantage of the SVM is that by using the functional forms of Eq.~(\ref{eq:svm_form}) in (\ref{eq:main_text_2}), we can perform the multivariable integration over $\mathbf{r_h}$, $\mathbf{r_e}$, and $\mathbf{r_{\ell}}$ analytically. That is, $M_{\lambda}(\mathbf{K},\mathbf{q})$ becomes a discrete sum over elements that are expressed in terms of the variational parameters. Given that it is sufficient to use a few tens of correlated Gaussians to accurately describe the electron and exciton states, and a few hundreds for the trion state, the calculation of  $M_{E_2'}(\mathbf{K},\mathbf{q})$ is efficient and fast. 

$\\$
The parameter values we use in the simulations for $A_u$,  $m_e$, $m_h$, $M_x$, $M_m$, $Z_{E_2'}$, $E_{E_2'}$, $\epsilon_{\text{SiO}_2}$, $d$, and  $\ell_{\pm}$ are listed in Sec.~\ref{sec:parameters} ($\epsilon_{\text{SiO}_2}$, $d$ and $\ell_{\pm}$ are parameters of $\epsilon_{3\chi}(q)$). 

\section{Derivation of Eq. (3) in the main text}
We assume that the extrinsic defects are located at a certain distance from the ML rather than distributed uniformly in the substrate. This assumption simplifies the analysis while not affecting the results qualitatively. The induced number of localized electrons in the ML is $N_i = n_dA$, where $n_d$ is their density and $A$ is the area of the ML.  Using second-order perturbation theory, the phonon-assisted recombination next to localization centers reads
\begin{eqnarray}
\frac{1}{\tau_{\lambda}(\mathbf{K})} &=& n_dA \frac{2\pi}{\hbar} \sum_{j,\mathbf{q}_t,\mathbf{q}_n}  \left| \frac{ P_{T\ell}(j,\mathbf{q}_t) M_{\lambda}(\mathbf{K},\mathbf{q}_n) }{  E_T + E_{\lambda} - [E_{\ell} + E_X +E_K]  } \right|^2 \delta \left( E_{j}(\mathbf{q}_t) + E_{\lambda}  - [ E_g - |E_X| + E_K]  \right) . \label{eq:2nd}
\end{eqnarray}
The photon quantum numbers are its three-dimensional wavevector $\mathbf{q}_t$ and polarization $j$. Its energy is $E_{j}(\mathbf{q}_t)=\hbar c q_t$ where $c$ is the speed of light. The phonon quantum numbers are its two-dimensional wavevector $\mathbf{q}_n$ and mode $\lambda$. Its energy, $E_{\lambda}(\mathbf{q}_n) = E_{\lambda}$, is assumed to be $\mathbf{q}$-independent due to the dispersionless nature of long-wavelength optical phonons. $E_T$, $E_{\ell}$ and $E_X$ are the energies of the localized trion, localized electron, and delocalized exciton, respectively. $E_g$ is the band-gap energy. Finally, the matrix elements in the numerator are due to the phonon-assisted capture process, $M_{\lambda}(\mathbf{K},\mathbf{q}_n)$, and photon emission, $P_{T\ell}(j,\mathbf{q}_t)$. The former follows from Eq.~(\ref{eq:main_text_2}), and the latter reads
\begin{eqnarray}
P_{T\ell}(j,\mathbf{q}_t) =  \langle  \Psi_0 \Psi_{\ell},\,  \{\mathbf{q}_t,j\}  | H_{\text{LM}} |  \Psi_T,\,\{0\}  \rangle \,,
\end{eqnarray}
where $H_{\text{LM}}$ is the light-matter interaction. The terms in curly brackets denote the photon state, and $\Psi_0\Psi_{\ell}$ denotes a state with one localized electron and a fully occupied valence band without delocalized electrons in the conduction band.  We rewrite Eq.~(\ref{eq:2nd}) using the fact that only $M_{\lambda}(\mathbf{K},\mathbf{q}_n)$ depends on $\mathbf{q}_n$,   
\begin{eqnarray}
\frac{1}{\tau_{\lambda}(\mathbf{K})} &=&  \left[ \frac{ n_dA \sum_{ \mathbf{q}_n}   \left| M_{\lambda}(\mathbf{K},\mathbf{q}_n) \right |^2  }{ (\Delta E - E_K )^2 } \right] \frac{ 1}{\tau_{\ell}} 
\end{eqnarray}
where $\Delta E = E_T + E_{\lambda} - [E_{\ell} + E_X]$, and
\begin{eqnarray}
\frac{ 1}{\tau_{\ell}} &=&  \frac{2\pi}{\hbar} \sum_{j,\mathbf{q}_t}  \left|  P_{T\ell}(j,\mathbf{q}_t) \right|^2  \delta \left( E_{j}(\mathbf{q}_t) + E_{\lambda}  - [ E_g - |E_X| ]  \right) \,,
\end{eqnarray}
is the trion radiative decay time. We have omitted the exciton kinetic energy, $E_K$, from the delta function  because of its negligible value: It is of the order of  1~meV at 5~K whereas all other energy scales in the delta-function are tens of meV (phonon energy), hundreds of meV (exciton binding energy) or $\sim$2~eV (band gap energy and photon energy).

\section{A compiled list of parameters we use in the simulations} \label{sec:parameters}

Figsurs 2 and 3 of the main text include simulation results for ML-WSe$_2$. The parameters of this system are

\begin{enumerate}
\vspace{-1mm}
\item The electron and hole effective masses are $m_e=0.29m_0$ and $m_h=0.36m_0$, respectively, where $m_0$ is the free electron mass. These results are taken from DFT calculations  \cite{Kormanyos_2DMater15}.
\vspace{-1mm}
\item We assume that the ML is supported on SiO$_2$ and exposed to air: $\epsilon_b=3.9$ and $\epsilon_b=1$.
\vspace{-1mm}
\item The thickness of the ML is $d=0.6$~nm. It is half the out-of-plane lattice constant in bulk TMDs, taking into account the distance between the chalcogen atomic sheets ($\sim$0.3~nm) and the van der Waals gap regions above and below the ML.
\vspace{-1mm}
\item The screening parameters of the ML, $\ell_{\pm}$, are the only fitting parameters in the simulations chosen to match the experimental results for the binding energies of trions and excitons. We have used $\ell_+=\ell_-=5.3d$ \cite{VanTuan_arXiv18}.
\vspace{-1mm}
\item The area of the unit cell is $A_u = \sqrt{3}a_{lc}^2/2 = 8.87~\AA^2$ where  $a_{lc}=3.2~\AA$ is the triangular lattice constant. 
\vspace{-1mm}
\item The atomic masses of tungsten and selenium are $M_{\text{W}}=M_m=3.05\cdot 10^{-22}$~g and $M_{\text{Se}}=M_x=1.31\cdot 10^{-22}$~g.
\vspace{-1mm}
\item The optical-phonon energy is $E_{E_2'}=30$~meV.
\vspace{-1mm}
\item The Born effective charge is $Z_{E_2'}=-1.16$, following DFT calculations \cite{Sohier_PRB16}. This value is in the correct ballpark since it yields the correct electron and hole mobilities at elevated temperatures (where the transport is limited by the Fr\"{o}hlich interaction). 
\vspace{-1mm}
\end{enumerate}
\vspace{-1mm}

%%%%%%%%%%%%%%%%%%%%%%%%%%%%%%%%

\end{widetext}

\end{document}